\documentclass[superscriptaddress,floatfix,aps,pre,twocolumn,notitlepage,shortbibliography,nofootinbib]{revtex4-1}
\usepackage[utf8]{inputenc}
\usepackage{amsmath}
\usepackage{amssymb}
\usepackage{refcount}
\usepackage{siunitx}
\usepackage{graphicx}
\usepackage{hyperref}
\usepackage[capitalise]{cleveref}
\crefname{equation}{}{}
\usepackage{microtype}
\usepackage{bm}
\usepackage[english]{babel}
\usepackage{color}

\newcommand{\comm}[2]{\left[ #1 , #2 \right]}
\newcommand{\acomm}[2]{\left\{ #1, #2 \right\}}

\newcommand{\br}{\mathbf{r}}
\newcommand{\hbr}{\hat{\mathbf{r}}}

\newcommand{\bp}{\mathbf{p}}
\newcommand{\hbp}{\hat{\mathbf{p}}}
\newcommand{\bE}{\mathbf{E}}
\newcommand{\bB}{\mathbf{B}}
\newcommand{\bA}{\mathbf{A}}
\newcommand{\bsigma}{\bm{\sigma}}

\newcommand{\hpi}{\hat{\pi}}
\newcommand{\hbpi}{\hat{\bm\pi}}
\newcommand{\balpha}{\bm\alpha}
\newcommand{\hF}{\hat{\mathcal{F}}}
\newcommand{\hH}{\hat{H}}
\newcommand{\hT}{\hat{T}}
\newcommand{\hepsilon}{\hat{\epsilon}}

\newcommand{\hO}{\hat{\mathcal{O}}}
\newcommand{\hEps}{\hat{\mathcal{E}}}
\newcommand{\hbSig}{\hat{\bm{\Sigma}}}

\newcommand{\B}{\textbf{B}}

\begin{document}
\author{Haidar al-Naseri}
\email{haidar.al-naseri@umu.se}
\author{Jens Zamanian}
\email{jens.zamanian@umu.se}
\affiliation{Department of Physics, Ume{\aa} University, SE--901 87 Ume{\aa}, Sweden}
\author{Robin Ekman}
\email{robin.ekman@plymouth.ac.uk}
\affiliation{Centre for Mathematical Sciences, University of Plymouth, Plymouth, PL4 8AA, UK}
\author{Gert Brodin}
\email{gert.brodin@umu.se}
\affiliation{Department of Physics, Ume{\aa} University, SE--901 87 Ume{\aa}, Sweden}
\title{Kinetic theory for spin-1/2 particles in ultra-strong magnetic fields}
\begin{abstract}
    When the Zeeman energy approaches the characteristic kinetic energy of
    electrons, Landau quantization becomes important.
    In the vicinity of magnetars, the Zeeman energy can even be relativistic.
    We start from the Dirac
    equation and derive a kinetic equation for electrons, focusing on the
    phenomenon of Landau quantization in such ultra-strong but constant
    magnetic fields, neglecting short-scale quantum phenomena. It turns out
    that the usual relativistic gamma factor of the Vlasov equation is
    replaced by an energy operator, depending on the spin state, and also
    containing momentum derivatives. Furthermore, we show that the energy
    eigenstates in a magnetic field can be computed as eigenfunctions of
    this operator. The dispersion relation for electrostatic waves in a
    plasma is computed, and the significance of our results is discussed.
\end{abstract}
\maketitle

\section{Introduction}
Quantum kinetic descriptions of plasmas are typically of most interest for
high densities and modest temperature~\cite{haas2011quantum}.
For this purpose, typically the the starting point is the Wigner-Moyal equation ~\cite{haas2011quantum,RevModPhys.83.885,daligault2014landau,materdey2003quantum,yang1991kinetic,misra2017}, or various
generalizations thereof also accounting for physics associated with the electron
spin, such as the magnetic dipole force~\cite{ZamanianNJP,andreev2017kinetic},
spin magnetization~\cite{ZamanianNJP,andreev2018separated} and spin-orbit
interaction~\cite{andreev2018separated}. Both weakly~\cite{asenjo2012semi,manfredi2019phase} and strongly~\cite{PhysRevE.96.023207,PhysRevE.100.023201}
relativistic treatments have been presented in the recent literature.

Certain quantum phenomena depend strongly on the magnitude of the
electromagnetic field, however, rather than on the density and temperature
parameters. Phenomena such as radiation reaction (reviewed in, e.g., Ref.~\cite{burton2014aspects}) and pair-creation
fall into this category.%
\footnote{While the classical radiation reaction, where the response is a smooth
function of the orbit, is a useful approximation as long as the emitted
spectrum is soft (dominated by photons with energies well below $mc^2$),
the description based on QED (see e.g. \cite{ilderton2013radiation,PhysRevX.8.011020}), containing discrete
probabilistic contributions due to the emission of high energy quanta,
is more generally applicable.}
Another field-dependent phenomenon is Landau
quantization~\cite{eliezer2005effects}, which becomes prominent whenever the Zeeman energy due to the
magnetic field is comparable to or larger than the thermal energy, or
the Fermi energy, in the case of degenerate electrons.

In the atmospheres of pulsars and magnetars~\cite{harding2006physics}, the electron motion may become
relativistic, and the magnetic field strength can be ultra-strong, i.e., the
Zeeman energy may be comparable to or even larger than the electron rest mass
energy~\cite{uzdensky}. Further information about pulsar properties
can be gained through the emission profiles, see, e.g., Refs~\cite{stinebring1984pulsar,rankin1990toward,han1998circular}.
Theoretical studies of wave propagation relevant for strongly magnetized objects have been made both with kinetic~\cite{melrose2009response} and hydrodynamic~\cite{asenjo2011hydrodynamical,zamanian20102} models.
However, most previous theoretical studies starting from the Dirac equation (see also~\cite{PhysRevE.96.023207,manfredi2019phase}) have been limited to cases where the
magnetic field strength is well below the critical field $B_{cr}=m^{2}c^{2}/\left\vert q\right\vert \hbar $.

Our objective in this work is to derive a fully relativistic kinetic model
of spin-1/2 particles, applicable for ultra-strong magnetic fields, i.e.
with \begin{equation}
    B\sim B_{cr}=m^{2}c^{3}/\left\vert q\right\vert \hbar,
\end{equation}
relaxing the
conditions given in previous works. Assuming that the electric field is low
enough to avoid pair creation (i.e., below the critical electric field $E_{cr}=m^{2}c^{3}/\left\vert q\right\vert \hbar $), we can use the
Foldy-Wouthuysen transformation~\cite{foldy1950,silenko2008} to separate particle states
from antiparticle states in the Dirac equation. Moreover, we limit ourselves
to the case where the characteristic spatial scale length of the fields is
much longer than the Compton length $L_{c}=\hbar /mc$.

Making a Wigner transformation of the density matrix, our approach results in an evolution
equation for a $2\times 2$ Wigner matrix, where the four components encode
information regarding the spin states.
However, the off-diagonal elements of the matrix is associated with the
spin-precession dynamics which is too rapid to be resolved by the
theory. Thus a further reduction is made, where only the diagonal
components representing the spin-up and the spin-down states relative to
the magnetic field remain.

Together with Maxwell's equations, we obtain a closed system describing the plasma
dynamics. A novel feature of the model is the energy expression, where the
usual gamma factor from classical relativistic theory is replaced by an
operator in phase space, also depending on the magnetic field.
Interestingly, the Landau-quantized states in a constant magnetic field turn
out to be eigenfunctions of the energy-operator of this theory, which is
helpful when computing the thermodynamic background state for the Wigner
matrix. To demonstrate the usefulness of the theory, we compute the
dispersion relation for Langmuir waves propagating parallel to an external
magnetic field. Finally, the consequences of the theory and applications to
astrophysics are discussed

\section{The strong Field Hamiltonian}
To derive our theory for ultra-strong magnetic fields, we will take the Dirac Hamiltonian
\begin{equation}
	\hH = \beta m + \hEps + \hO ,
	\label{DiracHamiltonian}
\end{equation}
as our starting point.
Here $m$ is the mass, $\hEps = q \phi ( \hbr )$, and $\hO = \balpha \cdot \hbpi$, and $\balpha$ and $\beta$ are the Dirac matrices.
Furthermore $\hbp$ is the canonical momentum, $\hbpi = \hbp - q \bA ( \hbr , t )$, $q$ is the charge, and $\phi$ and $\bA$ are, respectively, the scalar and vector potentials.
From now one we use units such that $ c = 1 $.

In their seminal paper Foldy and Wouthuysen~\cite{foldy1950} found a way to decouple the upper and lower two components of the four-spinor. 
For the general case, this can be done up to a given order in a suitably chosen expansion parameter.
In Ref.~\cite{foldy1950} this parameter was $ 1/m $, meaning that the expansion is valid for sufficiently small energies. 
In this paper we will use the results of Refs.~\cite{silenko2003,silenko2008} where a modified Foldy-Wouthuysen transformation was developed for the case where the expansion parameter is instead the scale-length of the fields.
This transformation hence makes it possible to take into account arbitrarily strong fields as long as we are only concerned with variations on sufficiently long scale-lengths.

The goal of a Foldy-Wouthuysen transformation is to obtain a Hamiltonian where the upper and lower pairs of components of the four-spinor are decoupled in the regime of interest.
An operator is called \emph{odd} if it couples the upper and the lower pairs of components of the four-spinor, and \emph{even} if it does not.
The odd and even terms of the Hamiltonian \cref{DiracHamiltonian} are $\hEps$ and $\hO$, respectively, satisfying $[\beta,\hEps] = 0$ and $\{\beta,\hO\} = 0$.

In~\cite{silenko2003}, Silenko found a unitary transformation of the Hamiltonian operator
\begin{equation}
    \label{UnitTransform}
    \hat{H}'= \hat{U}(\hat{H}-i\partial_t) \hat{U}^{\dagger} + i \partial_t,
\end{equation}
where
\begin{equation}
    \hat{U}^{(\dagger)}= \frac{\hepsilon + m \pm \beta\hO }{\sqrt{2\hepsilon(\hepsilon+m)  }},
\end{equation}
where $\hepsilon = \sqrt{m^2 + \hO^2 }$.
Furthermore, using the Dirac Hamiltonian $\hat{H}$ in \cref{UnitTransform}, one obtains
\begin{equation}
	\label{Silenko_Hamiltonian}
	\hat{H}'=\beta \hepsilon+ \hEps' + \hO',
\end{equation}
where
\begin{widetext}

\begin{align}
\hEps'&=
\hEps + \frac{1}{2\hT} \bigg(
            \comm{\hT}{\comm{\hT}{\beta \hepsilon + \hF }}
            - \comm{\hO}{\comm{\hO}{\hF}}
            - \comm{\hepsilon }{\comm{\hepsilon }{\hF}}
        \bigg)\frac{1}{\hT}\notag \\
  \hO'&=     
    \frac{1}{2\hT} \beta \bigg(
                \acomm{\hepsilon + m}{\comm{ \hO}{\hF}}
            -  \acomm{\hO}{\comm{\hepsilon }{\hF}}
        \bigg)\frac{1}{\hT}, 
\end{align}

\end{widetext}
where $\hT= \sqrt{2 \hepsilon (m + \hepsilon)}$ and $\hF = \hEps - i\hbar \partial_t$.
The Hamiltonian $H'$ still has odd terms, the anti-commutators in $\hO'$ are linear in $\hO$, i.e. linear in $\balpha$. However, the odd term $\hO'$ in $H'$ should be small compared to  $\hepsilon$, after choosing an expansion parameter.
Since we are interested in the effects of an ultra-strong magnetic field, our small expansion parameters will be the electric field strength $E/E_\text{cr}$(such that pair production is negligible) and the inverse scale-length of the fields $L_c/L$, but we will make no assumption on the \emph{magnetic} field strength.

Thus a second transformation can be preformed with the following operator
\begin{equation}
\label{SecondTransform}
	\hat{U}' = e^{iS'},  \quad \hat{S}' = -\frac{i}{4}\acomm{\hO'}{\frac{1}{\hepsilon}}.
\end{equation}
Since $\hO'$ is small, only the major corrections are  taken into account. Finally, the transformed Hamiltonian is
\begin{equation}
\label{EvenHamiltonian}
	\hH_{FW}= \beta \hat{ \epsilon} + \hEps' + \frac{1}{4} \acomm{ \hO'^2}{ \frac{1}{\hepsilon}}.
\end{equation}

This Hamiltonian has no odd terms, so we are done with the transformation of the Dirac Hamiltonian.  Until now, we have presented the Foldy-Wouthuysen transformation derived in Ref.~\cite{silenko2008}.  Our approach now is to calculate the commutators occurring in \cref{Silenko_Hamiltonian} to all orders in the magnetic field.
In doing this we will expand some of them in a series using Ref.~\cite{transtrum2005commutation}, see the Appendix for more details.
After the calculation, the Hamiltonian $H'$ in \cref{Silenko_Hamiltonian} becomes 
\begin{equation}
	\label{No_cum_Hamiltonian}
	\hat{H}' = \beta \hepsilon+ \hEps' + \hO'
\end{equation}
where
\begin{widetext}
    \begin{align}
        \hEps' = {}&
      \hEps+
        \frac{i m\mu_B}{\hT}
        \Bigg[
            i \hbar q
            \left(
                \frac{2}{\hT} + \frac{m}{\hepsilon \hT}
            \right)^2
            \left( \bE \times \bB \right) \cdot \hbpi
            - \frac{i \hbar q}{\hepsilon^2}
            \left( \bE \times \bB \right) \cdot \hbpi
            + i \hbSig \cdot ( \bE \times \hbpi - \hbpi \times \bE )
        \Bigg]
        \frac{1}{\hT} 
        \\
        \hO'
        = {}&
        \frac{im \mu_B \beta}{\hT}
        \Bigg[
            2  \balpha \cdot \bE \left( \hepsilon + m \right)
            - 2  \left( \frac{\hbpi}{\hepsilon} \cdot \bE \right) \balpha \cdot \hbpi
            +\frac{\hbar}{ \hepsilon}
            \left( \hbSig \cdot \partial_t \bB \right)
            \balpha \cdot \hbpi
        \Bigg]
        \frac{1}{\hT},
    \end{align}
\end{widetext}
where $\mu_B = q\hbar/2m$ is the Bohr magneton and
\begin{equation*}
    \hbSig
    =
    \begin{pmatrix}
        \bsigma & 0 \\
        0 & \bsigma
    \end{pmatrix}
    .
\end{equation*} 
Note that we kept all orders of the magnetic field. However we only kept up to first order of the combination of $\hbar$ with $E$  and $\nabla_r$.
In the expression for $\hH'$ there are still some odd operators, but since $\hO'$ is linear in $\mu_B E$ and $ \mu_B \partial_t B$,  they should be smaller than $\hepsilon$. Thus it is fine to only include the minor correction of the second transformation in \cref{SecondTransform}. Thus the Hamiltonian $\hat{H}_{FW}$ after the second transformation will have the same structure as in \cref{EvenHamiltonian}. However,
the anti-commutator in $  \hat{H}_{FW}$ in \cref{EvenHamiltonian} is proportional to the square of $\hO'$. Since we only kept up to first order of $E$ and $\partial_t B$,  we keep only the first two terms of $\hat{H}_{FW}$
 
\begin{widetext}
    \begin{equation}
        \label{Spin orbit Hamiltonian}
        \hat{H}_{FW}
        =
        \beta \hepsilon
        + q\phi(\hat{\textbf{r}})
        + \frac{\mu_B m }{\sqrt{2 \hepsilon(\hepsilon+m)}}
        \left(
            \hbSig \cdot (\hbpi \times \bE -  \bE \times \hbpi)
            - \frac{2\mu_B m}{\hepsilon^2} \left(
                1 + \frac{m^2}{2\hepsilon(m+\hepsilon) }
            \right)
            (\hbpi \times \bB) \cdot \bE
        \right) \frac{1}{\sqrt{2\hepsilon(\hepsilon + m)}},
    \end{equation}
\end{widetext}
where $\hepsilon = \sqrt{m^2 + \hbpi^2 - 2\mu_B m \hbSig \cdot \bB}$.
Taking the limit of weak $B$-field, i.e. keeping up to first order in $\mu_{B}B/m$, we recover the Hamiltonian in~\cite{silenko2008}.
The Hamiltonian in \cref{Spin orbit Hamiltonian}  includes the spin orbit interaction, see~\cite{PhysRevE.96.023207} for more details, but since $E \ll B$, it is negligible compared to the magnetic interaction.
We will therefore neglect the spin-orbit interaction from now on, as the main idea of this paper is to study the effects of a strong magnetic field on the dynamics of a plasma.
The transformed Hamiltonian is now
\begin{equation}
    \label{Final Hamiltonian}
    \hH_{FW}= \beta \sqrt{m^2+ \hbpi^2 - 2\mu_Bm \hbSig \cdot \bB}+ q\phi(\hat{x}).
\end{equation}
This Hamiltonian includes all orders of $\mu_B B/m$ and is fully relativistic,  hence is suitable to be used in deriving a kinetic equation for plasma in an environment where the magnetic field is of the order of the critical field $B_\text{cr} $.  
Note that all operators in both \cref{Spin orbit Hamiltonian} and \cref{Final Hamiltonian} are even, thus we will let $\beta \rightarrow 1$ and $\hbSig \rightarrow \bsigma$ from now on.

\section{Gauge-Invariant Wigner function}
\label{sec:gauge-invariant-wigner}
Now we want to derive a kinetic equation using the Hamiltonian \cref{Final Hamiltonian}. To do that, we start with the evolution equation for the density matrix $\hat{\rho}_{\alpha\beta}$ which is given by the Von Neumann equation
\begin{equation}
    \label{Von}
    i\hbar \partial_t \hat{\rho}_{\alpha \beta} = \comm{\hH_{FW}}{\hat{\rho}_{\alpha \beta} }.
\end{equation}
Our goal now is to transform this equation into a kinetic equation for the Wigner quasi-distribution function. Considering the gauge-invariant Wigner function derived by Stratonovich~\cite{stratonovich1956gauge}
\begin{widetext}
    \begin{equation}
        W_{\alpha\beta}(\textbf{r},\textbf{p},t)
        =
        \Big(\frac{1}{2\pi \hbar}\Big)^3 \int d^3\lambda
        \exp \left\{
            \frac{i \bm\lambda}{\hbar} \cdot 
			\left[
                \textbf{p} + q\int^{1/2}_{1/2} d\tau 
				\textbf{A} \left( \br + \tau \bm \lambda \right)
			\right]
		\right\}
		\rho_{\alpha\beta}\left(
            \br + \frac{\bm \lambda}{2}, \br - \frac{\bm \lambda}{2}, t
		\right).
    \end{equation}
    We express \cref{Von} in terms of $\rho(\br,\br')$ and use the identities~\cite{stratonovich1956gauge}
    \begin{align}
		F \left[ -i \hbar \nabla_r - q \bA(\br) \right] 
			\rho_{\alpha \beta}(\br,\br')
        & \rightarrow
        F\left[\bp - i\hbar/2 \nabla_r + im\mu_B \B\times \nabla_p  \right] W_{\alpha \beta}(\br,\bp)
        \\
		F \left[i \hbar \nabla_{r'} - q \bA(\br) \right] 
			\rho_{\alpha \beta}(\br,\br')
        & \rightarrow
        F\Big[\bp + i\hbar/2 \nabla_{r} - im\mu_B \B\times \nabla_p  \Big]W_{\alpha \beta}(\br,\bp).
    \end{align}
\end{widetext}
These identities can be utilized for a function $F$ that only depends on $\hbpi$. However, our Hamiltonian in \cref{Final Hamiltonian} depends on both $\bB(\hbr)$ and $\hbpi$, thus we divide the magnetic field as
\begin{equation}
    \bB(\hbr) = \B_0 + \delta \bB(\hbr),
\end{equation}
where $\bB_0$ is a constant strong magnetic field and $\delta\bB(\hbr)$ is a varying magnetic field.
Furthermore, a Taylor series around $\bB=\bB_0$ can be done, see Ref.~\cite{kumar1965} for more details.
However, the Taylor series gets more complicated for the higher order terms, thus a restriction on $\delta \bB(\hbr)$ needs to be done. Considering the case where $\mu_B \delta B/m \ll 1$, note that we did consider $\mu_BE/m\ll 1$ in the previous section, the Hamiltonian becomes
\begin{equation}
    \hH_{FW}= \beta \sqrt{m^2+ \hbpi^2 - 2\mu_B m \bsigma \cdot \bB_0} 
	+ q\phi(\hbr).
\end{equation}
While this Hamiltonian does not contain $\delta \bB$ explicitly, as will be seen below, the perturbed magnetic field will still be contained in the Lorentz force. 
Using this Hamiltonian and  keeping up to first order in $\nabla_r$ as we did in the derivation of the Hamiltonian,  the kinetic equation is
\begin{widetext}
    \begin{equation}
        \label{Kinetic_model}
        \partial_t W_{\alpha \beta}
        + \frac{1}{\epsilon'} \bp \cdot \nabla_r W_{\alpha \beta}
		+ q \left(
            \bE + \frac{1}{\epsilon'} \bp \times \bB
		\right)
            \cdot \nabla_p W_{\alpha \beta}
        = 0,
    \end{equation}
    where
    \begin{equation}
        \epsilon' = \sqrt{m^2+ \bp^2 - 2m\mu_B\bsigma \cdot \bB_0  - m^2\mu_B^2 (\bB_0\times \nabla_p)^2}.
    \end{equation}
\end{widetext}
Note that $\epsilon'$ is  a function of  $\bsigma$ and that the momentum derivatives act on everything to the right of the operator.
Thus, in the second and fourth terms of \cref{Kinetic_model} $ 1 / \epsilon' $ acts also on $ \bp $.
In order to get a scalar theory, we can Taylor-expand $\epsilon'$ around $\bsigma$
\begin{equation}
    \label{Taylor expanding Pauli}
    \epsilon'
    =
      \frac{1}{2} \Big(\epsilon'_{+} + \epsilon'_{-} \Big) I
    + \frac{1}{2} \Big(\epsilon'_{+} - \epsilon'_{-} \Big) \sigma_z
    ,
\end{equation}
where $I$ is the identity matrix and 
\begin{equation}
    \epsilon'_\pm = \sqrt{m^2+ \bp^2  \mp 2m \mu_B B_0  - m^2\mu_B^2 (\bB_0\times \nabla_p)^2}.
\end{equation}
Next, we note that if initially $W_{\alpha\beta}$ has no off-diagonal elements, let us say that $W_{11}=W_{+}$,  $W_{22}=W_{-}$, the evolution \cref{Kinetic_model} for $W_{+}$ and $W_{-}$ will decouple into separate equations for the spin-up and spin-down populations, as defined relative to $\bB_0$. 
While limiting ourselves to such initial conditions may seem unwarranted, the only thing left out by this restriction is the spin precession dynamics. 
However, since the time-scale for spin-precession is the inverse Compton frequency, we note that the present theory, based on the assumption $\partial/\partial t\ll c/L_c$, is not designed to resolve the spin precession dynamics anyway. 
Hence, from now on, we will be using the above representation for $W_{\alpha \beta}$, in which case \cref{Kinetic_model} decouples into the scalar equations for $W_{+}$ and $W_{-}$ as follows:
\begin{align}
    \label{Scalar_kinetic_model}
    \partial_t W_{\pm}
    + \frac{1}{\epsilon'_{\pm}} \bp \cdot \nabla_r W_{\pm}
    + q \Big[
        \bE + \frac{1}{\epsilon'_{\pm}} \bp \times \bB
    \Big] \cdot \nabla_p W_{\pm}
    & = 0.
\end{align}
The kinetic equation in \cref{Scalar_kinetic_model} is our main result in this work.
The new effects of this kinetic equation are hiding in $\epsilon '_\pm$. Firstly, we have all orders of the spin magnetic moment, compared to previous models~\cite{PhysRevE.96.023207,manfredi2019phase} where only the first order correction is included. Moreover, we have momentum derivatives in $\epsilon'$, which turn to be energy operators, see \cref{Background_section} for more details.

\Cref{Scalar_kinetic_model} describes the dynamics of an ensemble of spin-1/2 particles in an ultra-strong magnetic field in the mean-field approximation. 
In this approximation, the electric and magnetic fields are generated by the sources via
\begin{equation}
	\nabla \cdot \bE  = \rho_f , \quad \text{and} \quad 
		\nabla \times \bB = \mathbf{j}_f + \partial_t \bE ,
\end{equation}
where $\rho_f$ and $j_f$ are the free charge and current density respectively
\begin{align}
	\rho_f &= q\sum_{\pm} \int d^3 p \,  W_{\pm} \\
	\mathbf{j}_f &= q\sum_{\pm} \int d^3 p \, \frac{1}{\epsilon'_{\pm}}\, \bp
W_{\pm}.
\end{align}
Under the assumptions we have made, the bound sources arising from the spin are negligible, but these can be and have been included in other models, e.g. Refs.~\cite{ZamanianNJP,manfredi2019phase}

\section{Conservation laws}
To check the validity of the derived model, we derive the conservation law of energy and the mass continuity.  Starting with the mass continuity,
the number density of spin-up (spin-down) particles $n_\pm$ can be given by $n_\pm = \int d^3p \,  W_{\pm}$. To show that this quantity is conserved we take the time derivative of it and use \cref{Scalar_kinetic_model}. Since $\epsilon'$ is independent of $\br$, it is trivial to show that
\begin{equation}
	\partial_t n_\pm + \nabla_r \cdot \int d^3p \frac{1}{\epsilon'_{\pm}} \,\bp W_{\pm}
 = 0.
\end{equation}
The number densities are separately conserved because transitions between spin-up and spin-down states require absorption or emission of quanta with energies on the order of $m$.

Moving to the conservation of energy, the total energy density is
\begin{equation}
    E_\text{tot} = \frac{1}{2} (E^2+ B^2) +\sum_{\pm} \int d^3p\, \epsilon'_{\pm}W_{\pm}.
\end{equation}
We want now to show that the energy is conserved, taking the time derivative of $E_\text{tot} $, and using Maxwell's equations together with the kinetic equation, we get
\begin{equation}
	\partial_t E_\text{tot}  + \nabla_r \cdot \mathbf{K},
\end{equation}
where $\mathbf{K}$ is the energy flux
\begin{equation}
    \mathbf{K} =\bE \times  \bB + \sum_{\pm} \int d^3p\, \bp W_{\pm}.
\end{equation}
This is precisely what one would expect: the Poynting vector for the fields, and the kinetic energy flux for the particles as required in a relativistic theory.
Since we have neglected polarization and magnetization, there is no Abraham-Minkowski dilemma in this model; see Ref.~\cite{PhysRevE.100.023201} and references therein for a related discussion.
\section{Background Wigner function in a constant magnetic field}

\label{Background_section}
In principle we can compute the time-indpendent solutions for $W$ in a
constant magnetic field $\bB = B_{0} \mathbf{\hat{z}}$ by solving the
Dirac-equation for this geometry, making a sum over different particle
states, and then perform the Foldy-Wouthuysen and Wigner transformations of
\cref{sec:gauge-invariant-wigner}.
Except for the Foldy-Wouthuysen transformation, this was done in a covariant approach in Ref.~\cite{sheng2018wigner}.
However, here we will take a shorter route to arrive at the
same results. Noting that for a constant magnetic field, both the Dirac
equation and the Pauli equation results in electrons obeying a quantum
harmonic oscillator equation, we can make a trivial generalization of the
Pauli case \citep{ZamanianNJP}. Both for the Pauli and the Dirac equations, the spatial
dependence of the wavefunction in Cartesian coordinates can be expressed as
a Hermite polynomial times a Gaussian function \cite{melrose1983} only the
energy eigenvalues for the Landau quantized states are different.
Specifically, applying the Dirac theory, the energy of the Landau quantized
states become%
\begin{equation}
    E_{n\pm}
    =
    m\sqrt{
        1 + (2n + 1 \pm 1) \frac{\hbar \omega_{ce}}{m} + \frac{p_z^2}{m^2}
    }
    \label{relativisticenergy}
\end{equation}
where $n=0, 1 , 2 , \ldots$ corresponds to the different Landau levels for the
perpendicular contribution to the kinetic energy, the index $\pm$
represents the contribution from the two spin states, and the term
proportional to $p_{z}^{2}$ gives the continuous dependence on the parallell
kinetic energy. Since the Pauli and Dirac equations for individual particle
states have the same spatial dependence for the wave function, we can adopt
the expression for the Wigner function from Ref.~\cite{ZamanianNJP} (based on the
Pauli equation) with some relatively minor adjustments.
\begin{enumerate}
    \item Contrary to Ref.~\cite{ZamanianNJP}, we have made no Q-transform to introduce an independent spin variable, and thus the spin-dependence of Ref.~\cite{ZamanianNJP} reduces to $W_\pm$.

    \item The Wigner function of Ref.~\cite{ZamanianNJP} must be expressed in terms of the momentum, i.e., $m(v_{x}^{2}+v_{y}^{2})/2 \rightarrow (p_{x}^{2}+p_{y}^{2})/2m$.
    \item The non-relativistic energy of Ref.~\cite{ZamanianNJP} is replaced by the relativistic expression \cref{relativisticenergy} of the Dirac theory.
    \item The normalization of the Wigner function must be adopted to fit the present case.
\end{enumerate}
With these changes, the background Wigner function $W_{\pm}^{TB}$ for the case of
electrons in thermodynamic equilibrium can be written%
\begin{equation}
    W_{\pm}^{TB}
    =
	\frac{n_{0 \pm}}{\left( 2 \pi \hbar \right)^3} \sum_{n}
	\frac{2(-1)^n \phi_n (p_{\perp})}{\exp \left[ (E_{n,\pm }-\mu_{c})/k_{B}T\right] + 1},
    \label{Wigner-thermo}
\end{equation}
where
\begin{equation}
    \phi_n(p_\perp)
    =
    \exp \left(
        -\frac{p_{\perp}^2}{m\hbar \omega_{ce}^2}
    \right)
    L_n\left( \frac{2p_{\perp}^2}{m\hbar \omega_{ce}^2}\right),
    \label{Eigen-func}
\end{equation}
$n_0=n_{0+}+n_{0-} =\int (W_{+T}+W_{-T}) d^3 p$ is the electron number density of the plasma, $\mu_{c}$ is the
chemical potential, $T$ is the temperature, and $L_n$ denotes the Laguerre
polynomial of order $n$.

That the factor $\phi_n(p_{\perp}) $ gives us the proper Wigner
function for the Landau quantized eigenstates can be confirmed by an
independent check. Since the expression \cref{Wigner-thermo} contains no
dependence on the azimuthal angle in momentum space, we can write
\begin{widetext}
    \begin{equation}
        \epsilon^{\prime}_{\pm}
        =
        m\sqrt{
            1 + p_\perp^2/m^2
            - \mu_B^2 B_{0}^{2}\left(
                \frac{\partial }{\partial p_\perp} + \frac{1}{p_\perp}
            \right) \frac{\partial }{\partial p_\perp}
            \mp \frac{2\mu_B B_0}{m} + \frac{p_z^2}{m^2}
        }
        \label{simplified expression}
    \end{equation}
\end{widetext}
when $\epsilon^{\prime }_{\pm}$ acts on $\phi_n(p_\perp)$.
Computing $\epsilon^{\prime }_{\pm}\phi_n(p_\perp)$ by
Taylor-expanding the square root to infinite order, using the properties of
the Laguerre polynomials, and then converting the sum back to a square-root,
it is straightforward to verify the relation
\begin{widetext}
    \begin{equation}    
        \label{energy_operator}
        \epsilon^{\prime}_{\pm} \phi_n(p_\perp) 
        =
        m\left(
            1 + (2n + 1 \pm 1) \frac{\hbar \omega_{ce}}{m}+\frac{p_{z}^{2}}{m^{2}}
        \right)^{1/2}
        \phi_n(p_\perp) 
    \end{equation}
\end{widetext}
where $\omega_{ce} = \frac{|qB_0|}{m}$ is the electron cyclotron frequency,
confirming that $\phi_n(p_\perp) $ generates the proper energy
eigenvalues for the perpendicular kinetic energy and the spin degrees of
freedom.

While \cref{Wigner-thermo} gives the thermodynamic equilibrium
expression $W_{\pm}^{TB}$, we note that the plasma background state is not
necessarily in thermodynamic equilibrium. Making use of the property \cref{energy_operator}, we note that the most general time-independent solution $W_{0\pm}$ to \cref{Scalar_kinetic_model} of physical significance can be written in the form
\begin{equation}
    W_{0\pm}
    =
    \sum_{n}
    g_{n, \pm}(p_z) (-1)^n \phi_n(p_\perp) 
    \label{Background-general}
\end{equation}
where $g_{n \pm}(p_z)$ is a function that is normalizable, but
otherwise arbitrary, and the number of particles in each Landau quantized
eigenstate $n_{n, \pm}$ obeys the condition
\begin{align}
    n_{n\pm}
     = &
    \int g_{n\pm}(p_z) (-1)^n \phi_n (p_\perp) \, d^3p
    \notag \\
     \Rightarrow &
    \notag \\
    n_{n\pm }
     = & \frac{(2\pi\hbar)^3}{2}   \int g_{n\pm }(p_{z})dp_{z}.  \label{Norm-condition}
\end{align}
Naturally, the expressions for $W_{0\pm}$ and $W_{\pm}^{TB}$ presented here are of
most significance for relativistically strong magnetic fields, when Landau
quantization is pronounced. As a consequence, the above formulas will reduce
to more well-known expressions when the limit $\hbar \omega_{ce}/m\ll 1$ is
taken. Specifically, \cref{Wigner-thermo} will become a
relativistically degenerate Fermi-Dirac distribution in case we let $T=0$
and $\mu_{c} = E_{F} \gg \hbar \omega_{ce}$, where $E_{F}$ is the Fermi
energy. Alternatively, for $k_{B}T \gg E_{F}$ and $k_{B}T\gg \hbar \omega_{ce} $, \cref{Wigner-thermo} reduces to a Synge-Juttner distribution.

\begin{figure}
    \centering
    \includegraphics[width=\columnwidth]{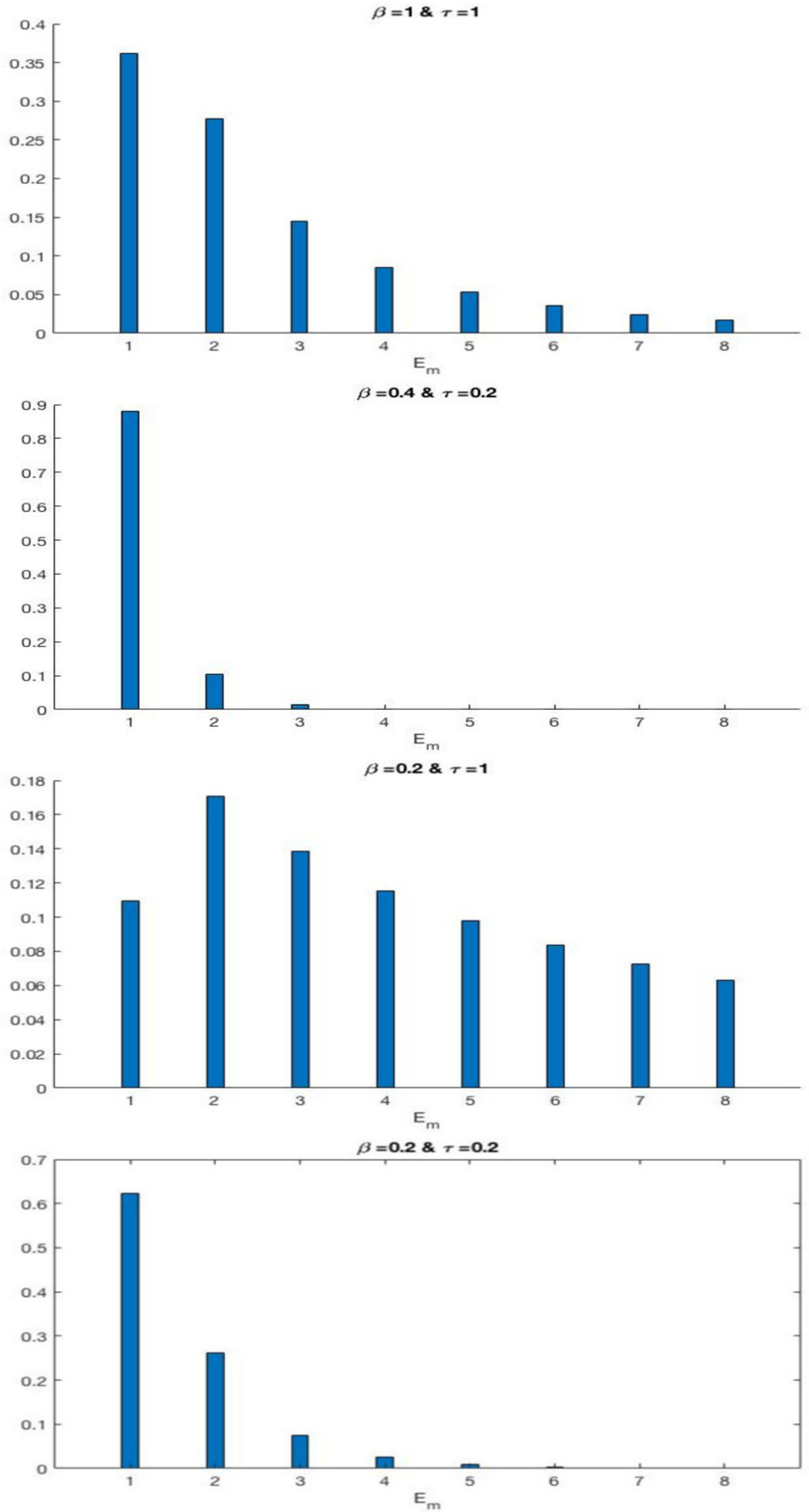}
    \caption{The normalized number density at different energy states $E_m$ for different values of the parameters $\beta=\mu_BB_0/m$ and $\tau= k_{BT}/m$.}
	\label{fig:background}
\end{figure}
To give a concrete illustration, in \cref{fig:background} we have made a bar chart for the normalized number density $n_{0n\pm}/n_0$ in the different energy states for a few values of the temperature and magnetic field, under the assumption that the density is low enough for the system to be non-degenerate, i.e. assuming $T\gg T_F$. 

As will be demonstrated in the next section, to a large degree the electrons behave as a multi-species system, where each particle species has its own rest mass, as given by \cref{relativisticenergy} but with $p_z=0$.
This is because the separation between Landau levels is on the order of the rest mass, and all excitation quanta with energies of that order have been neglected.
If we define the effective number density of each "species" (discrete energy-state) as
\begin{equation}
	n_{0n\pm} \equiv \frac{n_0}{\left(2\pi\hbar \right)^3} \int d^3p 
	\frac{2 \left( -1 \right)^n \phi_n ( p_\perp )}{ 
	\exp \left[ \left( E_{n\pm} - \mu_c \right) / k_BT \right] + 1},
\end{equation}
we see that $n_{0n\pm}$ essentially will be determined by the Boltzmann factors of \cref{Wigner-thermo}. However, for the cases where our model \cref{Scalar_kinetic_model} is of most interest, the magnetic field $B_0$ is strong enough to make relativistic Landau quantization prominent. Thus, in the next section and for the remainder of this paper, we will consider background distributions $W_{0\pm}$ where simplifications based on $\hbar \omega_{ce}/m\ll 1$ do not apply.

\section{Linear waves}

The operator in the square root in \cref{Scalar_kinetic_model} gives the impression that it is very technical and complex to apply the model in studying, e.g., waves in plasma.
In this section we consider electrostatic waves in a homogeneous plasma by using \cref{Scalar_kinetic_model}.
We consider the wave vector $\mathbf{k} = k \mathbf{e}_z$ and express the momentum $\mathbf{p}$ in cylindrical coordinates $p_\perp$, $\varphi_p$, and $p_z$.
To linearize \cref{Scalar_kinetic_model}, we separate variables according to $W_{\pm} = W_{0\pm}(p_\perp,p_z)+ W_{1\pm}(z, p_\perp, p_z, t)$, $\bE = E_1 \mathbf{e}_z $ and $\mathbf{B} =  B_0 \mathbf{e}_z$, where the subscripts $0$ and $1$ denote unperturbed and perturbed quantities respectively.
Moreover, the perturbed quantities follow the wave plane ansatz according to $W_{1\pm}=\tilde{W}_{1\pm} e^{i kz - i \omega t}$.
\Cref{Scalar_kinetic_model} is now
\begin{equation}
    \label{Linear Estat}
    \left(\omega-\frac{k p_z}{\epsilon'_{\pm}}\right) W_{1\pm}
    =
    - i qE_1 \frac{\partial W_{0\pm}}{\partial p_z}.
\end{equation}
The unperturbed Wigner function $W_{0\pm}$ is given by the thermal background Wigner function in \cref{Wigner-thermo}. Since the operators in $\epsilon'_{\pm}$ have been shown to be energy eigenvalues when acting on $W_{0\pm}$; we act on both sides by $\left( \omega-k p_z/\epsilon'_{\pm}\right)^{-1}$, such that
the perturbed distribution function becomes
\begin{equation}
    W_{1\pm}
    =
    \frac{-iqE}{\omega - kp_z/\epsilon'_{\pm}} \frac{\partial W_{0\pm}}{\partial p_z}
    =
    \sum_{n}  \frac{-iqE}{\omega - kp_z/E_{n\pm}} \frac{\partial W_{0n\pm}}{\partial p_z}
    ,
\end{equation}
where in the second equality we used that $W_{0\pm}$ is a sum of eigenfunctions of $\epsilon'_\pm$,
and the summation is over the Landau levels indexed by $n$.
Using Poisson's equation, the dielectric tensor for the electrostatic case is
\begin{equation}
    D(k,\omega)
    =
    1 + \frac{q^2}{k} \sum_{n, \pm}
    \int d^3p \, \frac{1}{\omega - kp_z/E_{n\pm}} \frac{\partial W_{0n\pm}}{\partial p_z}.
	\label{dispersionrelation}
\end{equation}
Note that if we set $\hbar$ to zero in $E_{n\pm}$ in the dielectric tensor, then the denominator in the the second term of the dielectric tensor is the same as for the relativistic Vlasov equation~\cite{alexandrov}.

The background distribution $ W_{0n\pm} $ will be divided into its eigenstates depending on the temperature and the magnetic field, see~\cref{fig:background}.
As a result, the dispersion relation \cref{dispersionrelation} is that of a relativistic multi-species plasma where each species has its own rest mass, $ E_{n\pm} $.

\section{Summary and Discussion}

In the present paper we have derived a kinetic model for plasmas immersed in a relativistically strong magnetic field, i.e., with a field strength of the order of the critical field.
Based on a Foldy-Woythausen transformation and a Wigner transformation, an evolution equation for the spin-up and spin-down components $W_\pm$ has been found. 
Besides having two components, the main difference to a classical relativistic model is that the gamma-factor in that theory is replaced by an operator containing momentum derivatives.
Since our theory is formulated in phase space, we stress that such operators are fundamentally different from the operators of Hilbert space, and to the best of our knowledge, this effect has not been seen in any previous quantum kinetic models.
An immediate effect of the energy being an operator, comes when studying the background Wigner function.
Classically, or in less advanced quantum mechanical models, the evolution equation does not predict the detailed momentum dependence for a given Landau level.
Thus the background expression has been put in by hand, and one has to return to the starting point of the theory (e.g., the Pauli or Dirac equations for single particles) to find proper expressions.
Here, however, the eigenvalue equation $ \varepsilon ^{\prime }W_{0}=E_{n,\pm }W_{0}$ determines the background state for a given Landau level (bar a constant for the number density), and there is no need to go outside the kinetic theory itself to find proper initial conditions.

To avoid some technical difficulties related to the operator orderings, we
have here divided the magnetic field into an ultra-strong but constant part
($B_{0}\mathbf{\hat{z}}$) and a fluctuating part $\delta \mathbf{B}$. This
approach allows for the treatment of large classes of problems in magnetar
atmospheres, for example linear and nonlinear wave propagating in
homogeneous backgrounds, even up to relativistic wave amplitudes, as long as
the flutuating part fulfills $\mu_{B}\left\vert \delta \mathbf{B}
\right\vert \ll m$.

Of course, a full modeling of the magnetar surroundings,
covering the dipole nature of the background source field, is beyond the
scope of such a theory. Moreover, in order to focus on the physics due to
ultra-strong magnetic fields, the present theory excludes effects such as
the magnetic dipole force, the spin magnetization, and the spin-orbit
interaction included in some previous models~\cite{asenjo2012semi,manfredi2019phase},
which can be justified for the long scale lengths and moderate frequencies that we focus on here.
In this context, however, it should be noted that omission of the spin-orbit
interaction is closely related to a correction term of the free current
density (see e.g. Eq.~(21) of Ref.~\cite{asenjo2012semi}). While the additional term
kept by Ref.~\cite{asenjo2012semi} is a small correction for the conditions studied
in this paper, it can contribute with currents perpendicular to $\mathbf{B}_0
$ that may be of importance for certain problems, specifically for
geometries where the (otherwise larger) parallel currents vanish. This and
other possible extensions of the current theory is a project for future
research.

To illustrate the usefulness of the present theory, we have computed the dispersion relation for Langmuir waves in a strong magnetic field for a relativistic temperature.
To a large extent, we find that the electrons behave as if they are divided into different species.
More concretely, each Landau level of the background plasma contributes to the susceptibility with a term similar to the classical relativistic expression, but with its own effective mass $m_{n\pm }=m\left( 1+(2n+1\pm 1)\hbar \omega_{ce}/m\right) ^{1/2}$.
We expect this result to generalize to some other problems, but not be completely general, as for certain problems, the difference between the standard gamma factor and the energy expression of the current theory will be apparent.
A more complete study of the effects due to ultra-strong magnetic fields is a project for future research.

\appendix

\section{Commutators}
The Hamiltonian in \cref{Silenko_Hamiltonian} contains some commutators of  functions of operators. To calculate these commutators,  we need to expand them in a series. We present here some of the calculations that were done in order to obtain the result in \cref{No_cum_Hamiltonian}.

Firstly, both $\hT$ and $\hepsilon$ are functions of $\hO$, since these functions can be expanded in a Taylor series of $\hO$, the commutator
\begin{equation}
\comm{\hO}{\hO^n}=0
\end{equation}
thus, we have that
\begin{equation}
\comm{\hT}{\hO}=\comm{\hepsilon}{\hO}=0.
\end{equation}
We can now rewrite \cref{Silenko_Hamiltonian} as
\begin{widetext}
    \begin{multline}\label{Hamiltonian_Ap}
        \hH'= \beta \hepsilon + \hEps + \frac{1}{2\hT} \bigg(
            \comm{\hT}{\comm{\hT}{ \hF }}
            -2\beta \comm{\hepsilon}{\hF}\hO
            +2\beta \comm{\hO}{\hF}\hepsilon
            +2m\beta \comm{\hO}{\hF}
            - \comm{\hepsilon }{\comm{\hepsilon }{\hF}}
            - \comm{\hO}{\comm{\hO}{\hF}}
        \bigg)\frac{1}{\hT}, 
    \end{multline}
\end{widetext}
Looking  firstly at the commutator of $\hepsilon$ and $\hF$  (the same result can be used to the commutator of $\hT$ and $\hF$), we have
\begin{equation}
\comm{\hepsilon}{\hF}= \comm{\hepsilon}{q \phi(\hbr)}+ i\hbar\partial_t \hepsilon.
\end{equation}
To calculate the commutator $\comm{\hepsilon}{\phi (\hbr)}$, we expand the functions in the commutators in a series~\cite{transtrum2005commutation}
\begin{align} \label{Serie_expansion_1}
	\comm{\hepsilon}{\phi(\hbr)}
	&=
	- \sum_{k=1}^{\infty} \frac{(i\hbar)^k }{k!} \hepsilon ^{(k)} \phi^k(\hbr)
	\notag
	\\
	& \approx -i\hbar \frac{\partial \hepsilon  }{\partial \hat{\pi}_i}\nabla_i \phi (\hbr),
\end{align}
where in the last equality,  higher derivative terms were neglected in accordance with the long scale approximation. Thus, we have
\begin{equation}\label{Help_function1}
    \comm{\hepsilon}{\hF}= i\hbar q \frac{\bm\hpi}{\hepsilon}\cdot \bm E
    .
\end{equation}

Next, we calculate the commutator
\begin{align}
\label{Commutator_1}
\comm{\hepsilon}{\comm{\hepsilon}{\hF}}&=  i\hbar q\frac{1}{\hepsilon}\comm{\hepsilon}{\bm \hpi\cdot \bm E} 	\notag \\
&= i\hbar q\frac{1}{\hepsilon} \Big(
\comm{\hepsilon}{\hat{\pi}_j }E_j +  \hat{\pi}_j \comm{\hepsilon}{E_j}
\Big) 	\notag \\
&=i\hbar q\frac{1}{\hepsilon} 
\comm{\hepsilon}{\hat{\pi}_j }E_j, 
\end{align} 
where in the last equality, we neglected the commutator of $\hepsilon$ and $E_j$ in accordance with the long-scale approximation.
For the the commutator of $\hepsilon$ and $\hat{\pi}_j$, we need to expand it in a series. However this time, we have a commutator of functions  that depend on both $\hat{p}$ and $\hat{r}$. Using the result from \cite{transtrum2005commutation}, we expand the commutator in a series 
\begin{align}
        \comm{\hepsilon}{\hat{\pi}_j }E_j
        &=
        \sum_{k=1}^{\infty} \frac{(i\hbar)^k }{k!}
        \bigg(
            \frac{\partial^k \hat{\pi}_j }{\partial^k r_i} \frac{\partial^k \hepsilon}{\partial^k p_i} 
            -\frac{\partial^k \hepsilon}{\partial^k r_i} \frac{\partial^k \hat{\pi}_j }{\partial^k p_i} 
        \bigg)E_j
        \notag
        \\
        & \approx
        i\hbar q\frac{\hbpi}{\hepsilon} \cdot (\textbf{E} \times \textbf{B}),
\end{align}
where in the last equality we only kept up to $k=1$ in the summation since higher order terms vanish in the long-scale approximation. Finally, we have
\begin{equation}\label{Help_function2}
\comm{\hepsilon}{\comm{\hepsilon}{\hF}}= ( i\hbar q)^2\frac{\hbpi}{\hepsilon^2} \cdot (\textbf{E} \times \textbf{B}).
\end{equation}
Note that $\comm{\hT}{\comm{\hT}{\hF}}$ is calculated in the same way
\begin{equation}\label{Help_function3}
\comm{\hT}{\comm{\hT}{\hF}}= (i\hbar q)^2 \bigg( \frac{1}{\hT}\Big(2+ \frac{m}{\hepsilon} \Big)
\bigg)^2
\hbpi \cdot (\textbf{E} \times \textbf{B}).
\end{equation}

Now we will calculate the commutator of $\hO$ and $\hF$
\begin{equation}\label{Help_function4}
\comm{\hO}{\hF}=i\hbar q \alpha \cdot \textbf{E}.
\end{equation}
We did not need to do any approximation in calculating this commutator since it is linear in $\hO$. Finally, we calculate
\begin{align}\label{Help_function5}
 \comm{\hO}{\comm{\hO}{\hF}}&= i\hbar q \Big( \alpha_i \hat{\pi}_i\alpha_jE_j- \alpha_jE_j\alpha_i\hat{\pi}_i\Big) 	\notag \\
 &= -\hbar q \hbSig \cdot \Big( \hbpi \times \textbf{E} - \textbf{E}\times \hbpi
 \Big),
\end{align}
where in the last equality we have used
\begin{equation*}
\alpha_i\alpha_j= \delta_{ij}+ i \varepsilon_{ijk} \Sigma_k.
\end{equation*}
Using \cref{Help_function1,Help_function2,Help_function3,Help_function4,Help_function5}
in \cref{Hamiltonian_Ap}, we get the Hamiltonian $\hat{H}'$ in \cref{No_cum_Hamiltonian}.
\bibliography{Strong_field}{}

\end{document}